\documentclass[aps, prl, twocolumn,showpacs]{revtex4}
\usepackage{graphicx}
\usepackage{dcolumn}
\usepackage{bm}
\usepackage{color}
\begin{document}

\title{The unusual electronic structure of the "pseudo-ladder"
compound CaCu$_2$O$_3$}

\author{T.K.\ Kim, H.\ Rosner$^{1}$\cite{helge}, S.-L.\ Drechsler, Z.\
Hu\cite{hu}, C.\ Sekar, G.\ Krabbes, J.\ M\'alek\cite{mal}, M.\
Knupfer, J.\ Fink, H.\ Eschrig}

\affiliation{ Leibniz-Institute for Solid State and Materials Research
Dresden, P.O.Box 270116, D-01171 Dresden, Germany\\
$^{1}$ Department of Physics, University of California, Davis, CA
95616, USA}

\date{\today}

\begin{abstract}
Experimental and theoretical studies of the unoccupied electronic
structure of CaCu$_2$O$_3$ single crystals have been performed using
polarization-dependent x-ray absorption spectroscopy and band
structure calculations.  The measured hole distribution shows an
unusual large number of holes in orbitals parallel to the interlayer
direction which is in agreement with the theoretical
analysis. CaCu$_2$O$_3$ deviates significantly from the standard
$pd\sigma$ cuprate picture. The corresponding strong interlayer
exchange is responsible for the missing spin gap generic for other
two-leg ladder cuprates.
\end{abstract}
\maketitle

The electronic structure and magnetic properties of cuprates
exhibiting various interesting physical properties have been
intensively studied in recent years.  In particular, considerable
attention has been attracted by so called
spin-ladders.\cite{Rice,Dagotto,Mueller} Their study might contribute
to a complete understanding of spin and charge excitations in
high-$T_c$ superconductors since in the presence of a spin gap
$d$-wave superconductivity might occur in spin-ladder
compounds.\cite{Rice}  Superconductivity under high pressure has been
reported in Sr$_{14}$Cu$_{24}$O$_{41}$ \cite{Uehara,Isobe}.  So far,
only a few studies were carried out for CaCu$_2$O$_3$ (belonging
structurally to the ladder family; see Fig.\ 1).
\vspace{-0cm}
\begin{figure}[b]
\includegraphics[angle=-90, width=7.5 cm]{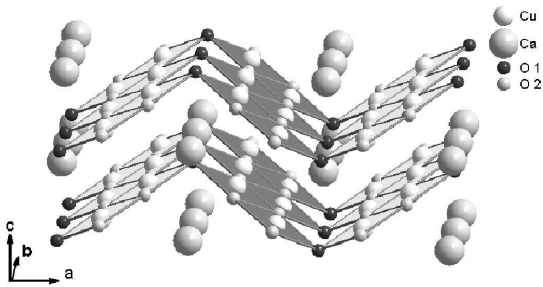}
\caption{\label{fig:structure} The crystal structure of CaCu$_2$O$_3$.
The corner-shared CuO$_2$ zigzag chains running 
along the $b$-axis are alternatingly 
tilted by nearly 28.55$^\circ$ forming positively and negatively buckled
ladders with ``kinked'' rungs in $a$-direction.}
\end{figure}
However, the magnetic susceptibility $\chi (T)$ is
\cite{Kiryukhin,Ruck} quite different from that of the
prototype two-leg ladder system SrCu$_2$O$_3$ \cite{Azuma}. From $\chi
(T)$ an antiferomagnetic ordering below a N\'eel temperature
$T_N\approx$25 K was concluded.  In order to contribute to the
understanding of ladder compounds in general and to clarify the
character of the hole distribution in CaCu$_2$O$_3$, polarization
dependent x-ray absorption spectroscopy (XAS) studies of single
crystalline samples were performed.  We report the experimentally
studied unoccupied electronic structure together with that derived
from band structure calculations within the local density
approximation (LDA) and show that at variance with usual ladder
compounds, the hole distribution in CaCu$_2$O$_3$ has a strong
non-planar nature.

The CaCu$_2$O$_3$ single crystals were grown by the traveling solvent
floating zone (TSFZ) method.  The as grown crystals were found to be
phase pure as confirmed by X-ray diffraction, energy dispersive X-ray
(EDX) and thermogravimetric analyses \cite{Sekar}.  Further, the EDX
results indicate a deficiency of Ca, with a balancing excess in Cu
corresponding to a nonstoichiometric composition
Ca$_{0.86}$Cu$_{2.14}$O$_{2.93}$ in agreement with the structure
refinement by Ruck {\it et al}.\cite{Ruck} The lattice constants were
determined to be
$\bm{a}$ = 9.946 \AA, 
$\bm{b}$ = 4.079 \AA {} and
$\bm{c}$ = 3.460 \AA {} 
in agreement with the values found in the literature.\cite{structure}
Single crystal samples with the size of $\approx$ 5$\times$5$\times$2
mm$^3$ were cut from the grown rod for the XAS measurements. The
determination of the sample orientation was carried out by taking
Laue-images (image plate), followed by a simulation of the surface
orientation.  The structure of this material, shown in
Fig.~\ref{fig:structure}, is similar to that of SrCu$_2$O$_3$
\cite{Hiroi}. It consists of an array of ladder-like structures with
quasi-one-dimensional copper-oxide chains extending along the
crystallographic $\bm{b}$ direction.  However, while in the Sr
compound the Cu-O-Cu bond angle in the rungs is
180\raisebox{1ex}{\scriptsize o}, in CaCu$_2$O$_3$ it turns out to be
$\approx$ 123\raisebox{1ex}{\scriptsize o}.

The XAS experiments were carried out using linearly polarized light
from the U49/1-PGM beam-line at the synchrotron light source BESSY II
in Berlin \cite{Jung}.  The energy resolution of the monochromator was
set to be 280 and 660 meV at the O 1$s$ and Cu 2$p$ absorption
thresholds, respectively.  A core-level excitation whose absorption
coefficient is small compared to the total absorption, as in the case
for the O 1$s$ edge, leads to a very poor signal-to-background ratio
in total electron yield (TEY), but has the distinct advantage that
self absorption effects in the fluorescence yield (FY) mode remain
small. The situation is just reversed in the case of the relatively
strong Cu 2$p$ absorption.  Therefore, for the O 1$s$ and the Cu 2$p$
absorption spectra we chose the FY and TEY detection mode,
respectively.  The fluorescence was detected by a solid state Ge
detector, placed at 45\raisebox{1ex}{\scriptsize o} angle with respect
to the incident photon beam. A total electronic yield was measured by
means of a sample current amplifier.  The sample was mounted on a
two-axis rotatable sample holder, which allowed us to measure spectra
with the light polarization vector $\bf{E}$ set parallel to one of the
crystallographic $\bm{a}$, $\bm{b}$ or $\bm{c}$ axes.  Energy
calibration was performed by comparison of the O 1$s$ and Cu 2$p$ XAS
signal of a CuO sample with corresponding electron energy-loss data
\cite{Nucker}.
In all cases the data were corrected for the energy-dependent incident
photon flux and, in the case of O 1$s$, for self-absorption effects
following a procedure described elsewhere \cite{Troger}.  The spectra
for different crystal orientations are normalized 60 eV above the
absorption threshold, where the final states are nearly isotropic
free-electron-like.  All measurements were performed at 300 K with the
vacuum in the chamber better then 10\raisebox{1ex}{-9} mbar during the
measurements.
\begin{figure}
\includegraphics[width=8 cm]{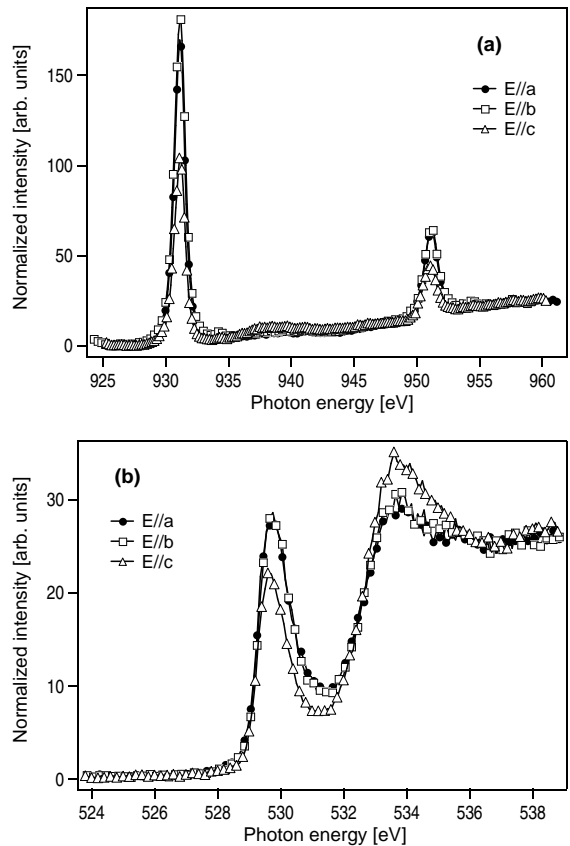}
\caption{\label{fig:XAS}  Polarization dependent XAS spectra at the (a) Cu
2$p$ and (b) O 1$s$ absorption edges of CaCu$_2$O$_3$.}
\end{figure}

Fig.\ \ref{fig:XAS}(a) shows the polarization dependent Cu 2$p$
XAS spectra of CaCu$_2$O$_3$ for $\bf{E}$ parallel to
all three crystallographic axes.  As transitions into Cu 4$s$
states are weak \cite{Teo}, these measurements probe mainly the hole
distribution in the Cu 3$d$ orbitals.  All spectra show a narrow
peak at 931.1 eV and 951.2 eV (the so called ``white line''), which
is associated with excitations from the Cu 2$p$$_{3/2}$ and Cu 2p$_{1/2}$
levels into Cu 3$d$ contributions to the upper Hubbard band
(UHB) \cite{Fink}.  While the data for 
$\bf{E} \| \bm{a,b}$ 
are very similar, a clear variation in intensity is found going to
$\bf{E} \| \bm{c}$.  
The white line is suppressed for
$\bf{E} \| \bm{c}$, 
but the features at 936-943 eV 
are higher in intensity compared to the other polarization directions.

The polarization dependent O 1$s$ XAS edges are
shown in Fig.~\ref{fig:XAS}(b).  These measurements probe empty
electronic states located in the 
O 2$p_{x}$ ( $\bf{E} \| \bm{a}$ ), 
O 2$p_{y}$ ( $\bf{E} \| \bm{b}$ ), and 
O 2$p_{z}$ ( $\bf{E} \| \bm{c}$ ) orbitals.  
The features directly above the absorption onset (i.e. below 531 eV)
can be assigned to transitions to the UHB resulting from hybridization
of O 2$p$ orbitals with Cu 3$d$ states.  The O 1$s$ absorption is
considerably different for $\bf{E} \| \bm{a}$, $\bm{b}$ and $\bf{E} \|
\bm{c}$.  For $\bf{E}\|\bm{c}$ the pre-edge peak intensity is lower
and the feature at 534 eV, which most likely arises from a combination
of O and Ca states, is more pronounced.  As for the Cu 2$p$ edge, for
the O 1$s$ absorption a relatively big fraction of holes was found in
$\bm{c}$ direction compared to the ``one-leg ladder'' compound
SrCuO$_2$ \cite{Knupfer} and the two-leg-ladder
Sr$_{14}$Cu$_{24}$O$_{41}$ \cite{Nucker-00}.  This unusual behavior
will be discussed below together with LDA results.

\begin{figure}
\includegraphics[angle=-90, width=8 cm]{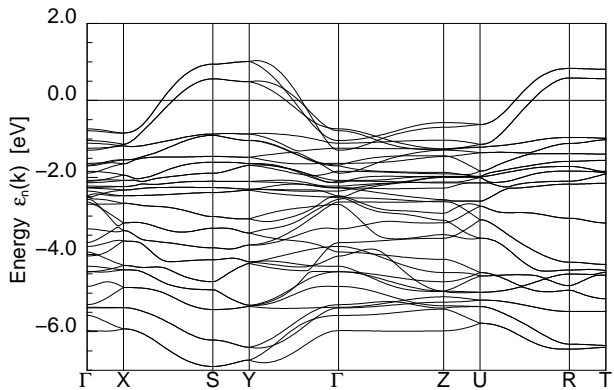}
\caption{\label{fig:LDA} The LDA band structure of CaCu$_2$O$_3$ 
near $E_F$.}
\end{figure}

In order to get insight into the projected density of states in the
different directions, band structure calculations were performed using
the full-potential nonorthogonal local-orbital minimum-basis scheme
\cite{Koepernik99} within the LDA. In the scalar relativistic
calculations we used the exchange and correlation potential of Perdew
and Zunger \cite{Perdew81}.  Cu ($3s$, $3p$, $4s$, $4p$, 3$d$),
O(2$s$, 2$p$, 3$d$), and Ca ($3s$, $3p$, $4s$, $4p$, 3$d$) states,
respectively, were chosen as the basis set. All lower lying states
were treated as core states. The inclusion of Cu and Ca (3$s$, 3$p$)
states in the valence states was necessary to account for
non-negligible core-core overlaps. The O 3$d$ states were taken into
account to increase the completeness of the basis set.  The spatial
extension of the basis orbitals, controlled by a confining potential
\cite{Eschrig89} $(r/r_0)^4$, was optimized to minimize the total
energy.  The results of the paramagnetic calculation (see
Fig.~\ref{fig:LDA}) show a valence band complex of $\approx$ 8 eV
width with four bands crossing the Fermi level ($E_F$=0) according to
the four copper atoms per unit cell. An analysis of partial densities
of states (not shown) shows that the valence band is mainly built by
Cu 3$d$ and O 2$p$ states with small contributions of Cu 4$s$ and 4$p$
states at the bottom of the valence band due to hybridization. The
contribution of Ca states is negligible.

The two band complexes at $E_F$ (see Fig.~\ref{fig:LDA}) have the
typical bandwidth of CuO$_3$-chain derived compounds of $\approx$ 2 eV
and are nearly half-filled \cite{Rosner}.  Furthermore, strong
correlation effects are present which explain the experimentally
observed insulating ground state. As one would expect, the main
dispersion of these two antibonding Cu 3$d$-O 2$p$ band complexes
(with respect to the Cu-O bonds) occurs along the $\Gamma$-Y
direction, corresponding to the crystallographic $\bm{b}$ direction of
the CuO$_2$ double chains (legs).  The upper band along ($\Gamma$-X,
X-S, S-Y, U-R, R-T, and the group with stronger dispersion along
$\Gamma$-Y is antibonding in nature whereas their counter part is
bonding (now with respect to the rung Cu-Cu bond of the ladder). They
are split by about 0.5 eV due to the interactions $t_{\perp}$ via the
rung.  The weak interladder coupling causes a further small splitting
along the $\Gamma$-X, $\Gamma$-Y, $\Gamma$-Z and Z-U lines.  Within
the corresponding zigzag representation the shift of the neighboring
leg by $b/2$ in real space results in a doubling of the Brillouin zone
(BZ) along the $b$-axis: $(-2\pi/b \leq k_y \leq 2\pi/b $).  Then that
secondary splitting can be understood as a band defolding in the
original twice as short BZ. Surprisingly, the ``interlayer''
dispersions along $\Gamma$-Z (crystallographically $\parallel$ to
$\bm{c}$) are quite strong for a cuprate compound being of the {\it
same} order as that along the rungs as mentioned above, i.e.\ of the
order of $\sim$ 500 meV. For the bonding bands the $2p_z$ O(1)
orbitals involved due to the strong buckling (which produce there
locally odd parity (under reflection along the $\bm{c} $-axis) of the
corresponding ``molecular'' rung orbitals) cause negative dispersion
along $\Gamma$-Z. Contrary the usual positive cosine-like dispersion
of the antibonding band is caused by predominant Cu $3d$ contributions
with even parity and a much smaller admixture of O 2$p_z$ states.  The
dispersion along $\bm{a}$ is rather small due to the weak coupling
between the sub-chains of the double (zigzag) chain (shaded structural
blocks in Fig.\ \ref{fig:structure}. The reason for the weak
inter-sub-chain coupling is the nearly 90$^\circ$ Cu-O-Cu bond between
neighboring Cu in different sub-chains.
The relative number of holes in the O2$p$ and Cu 3$d$ orbitals 
pointing along the respective crystal directions can be derived 
from the intensity of the lowest lying absorption feature in the 
O 1$s$ and Cu 2$p$ absorption edges, respectively.  
These numbers are listed in Table~\ref{tab:table1} and 
compared to those obtained for the projected density of states 
form our band structure calculations.

\begin{table}
\caption{\label{tab:table1} Relative number of holes for different 
polarization directions from XAS experiment and 
LDA calculations.}
\begin{ruledtabular}
\begin{tabular}{llll}
n$_h$ , \% &{\em\bf{a}}&{\em\bf{b}}&{\em\bf{c }}\\
\hline
O  $_{theory}$ & 35 & 44 & 21 \\
O  $_{exp}$ & 37 & 36 & 27 \\
Cu$_{theory}$ & 47 & 36 & 17 \\
Cu$_{exp}$ & 36 & 41 & 23 \\
\end{tabular}
\end{ruledtabular}
\end{table}

Table ~\ref{tab:table1} demonstrates that the XAS and the
theoretically derived numbers are in good agreement.  Both reveal that
about 40 \% of the holes occupy orbitals along the $\bm{a}$ and
$\bm{b}$ directions.  Strikingly, a relatively big portion ($\approx$
20 \%) of holes is found in orbitals along the $\bm{c}$ direction.  At
first glance this anomalous large number of non-planar holes as
compared to the ``normal ladder'' \cite{Nucker-00} could be understood
considering simply projections due to the buckled structure of
CaCu$_2$O$_3$.  From the measured relative O hole number 0.37 in
$a$-direction one has a local "in-plane (plaquette) of
0.37/$\cos^2\theta =$0.48 nominal hole concentration, where $\theta
\approx \pm 28.55^\circ $ is the local tilting angle (see Fig.\
1). Projecting this number on the $\bm{c}$-axis, a contribution of
only 0.11 nominal O holes would be expected in contrast with the
observed 0.27 nominal ones. Hence, there are about 0.16 nominal holes
from orbitals which usually do not enter the ground state of a local
CuO$_4^{-6}$ plaquette configuration. In order to illustrate the
absolute charge distribution, we adopt a typical Cu-O hole
distribution ratio with 2.4 Cu and 1.6 O holes per unit cell. Then one
has $\approx$0.26 absolute holes residing in nonstandard plaquette O
2$p$ orbitals. Within the LDA one arrives at $\approx$0.104 nominal
``anomalous'' O 2$p_z$ holes which reside mostly (71\% )in the O(1)
2$p_z$ states on the rung.  Remarkably, just these states yield an
important contribution to the coupling in $c$-direction as discussed
above for the bonding band.  While the Cu-O(1)-Cu bond angle
(180$^\circ -2\theta$) at the rungs decreases from 180$^{\circ}$, as
it would be in the case of a ``normal'' planar ladder (e.g.\
SrCu$_2$O$_3$), the coupling of the electronic states of two legs
within the ladder through the rung oxygen is reduced, and above some
critical angle $\theta=45^\circ -\delta$ in approaching the almost
"decoupled" chain limit at 45$^\circ$ it might become even smaller
than that in $\bm{c}$ direction.  Considering the Cu holes, we arrive
at $\approx$0.12(0.3) anomalous nominal (absolute) holes.  Thus in
total $\approx$0.56 (Cu + O) anomalous holes of 4 holes per unit cell
are moved into $\bm{c}$-axis oriented orbitals. This is much more than
usually observed in various planar cuprates ($<$ 0.05 holes per Cu
which is mainly attributed to the experimentally available slightly
nonideal polarization in the present day XAS experiments; see e.g.\
Refs.\ \onlinecite{Nucker-00,Neudert}).  This clear effect then also
leads to more isotropic spin interactions for CaCu$_2$O$_3$ which is
consistent with the observation of a 3D N\'eel like transition at
T$_{{N}}$= 25 K in magnetic susceptibility measurements
\cite{Kiryukhin,Ruck}, while SrCu$_2$O$_3$ stays in the spin gaped
nonmagnetic state down to $T=0$ \cite{Azuma}.  As our LDA-results
point to a comparable rung and interlayer hopping integral
$t_{\perp}\sim$ 250 meV and $t_z \approx \pm$ 125 meV, comparable
antiferromagnetic exchange integrals $J_{\perp} \sim $ 50 meV and $J_z
\sim$ 20 meV within a simple bilinear spin-1/2 Heisenberg model
picture can be expected.\cite{remspin} The consideration of a
ferromagnetic contribution of the same order as $J_z$ due to Hund's
rule coupling between O(1) 2$p_x$ and 2$p_z$ orbitals introduced by
the rung buckling \cite{rem2}, would bring both exchange integrals
still closer to each other. Then, the critical regime expected near
$J_z\sim 0.3\sqrt{J_{\parallel}J_{perp}}$ is arrived (compare Ref.\
\onlinecite{capriotti}).  In this way, the standard ladder picture
becomes invalid and we arrive at a picture of coupled ordered chains
whose ground state is a N\'eel-like state (period doubling along $b$-
and $c$-directions but an incommensurate ordering in
$\bm{a}$-direction due to the frustrated ladder arrangement).  The
numbers given above should be compared with those of the spin gaped
planar ladder compound SrCu$_2$O$_3$ \cite{Mueller} where at a
comparable large $J_{\parallel}\approx$ 130 to 150 meV, a larger
$J_{\perp}\approx $ 72 meV has been found.  But most importantly, only
a tiny $J_z \sim 1$ meV can be expected.\cite{remy} Thus,
CaCu$_2$O$_3$ is only a ``pseudo-ladder'' compound which differs
markedly from standard two-leg ladders.  Since the interladder
exchange along $\bm{a}$ is weak, we arrive at a picture which is
essentially in line with that put forward by Kiryukhin {\it et al.}
\cite{Kiryukhin} who suggested that CaCu$_2$O$_3$ can be described as
an array of 2D "$\bm{b}$-$\bm{c}$ double-planes" in which spins-1/2
are coupled by spatially anisotropic antiferromagnetic exchange
interactions described by the isotropic (in spin space) Heisenberg
model.

In conclusion, we have studied the unoccupied electronic structure of
the ``pseudo-ladder'' compound CaCu$_2$O$_3$.  The experimental XAS
data of the relative hole numbers is in good agreement with the LDA
values.  Further calculations to study correlation and impurity
effects are necessary to address some remaining small deviations.  At
variance with the normal ladder picture (and more generally also with
all other known cuprates comprised from isolated or shared
CuO$_4$-plaquettes in straight or planar networks) the hole
distribution has a less pronounced local O2$p$Cu$d\sigma $ character.
Here it is related to the missing spin gap and the
antiferromagnetically ordered N\'eel-like state below 25 K.  A broad
study of the properties of this unique cuprate, including doping, is
of interest for the ladder and the cuprate physics in general.  In
particular, any doping induced superconductivity would be different
from the spin-gap based scenario proposed for two-leg ladder cuprates.

\begin{acknowledgments}
Support by the DFG (Fi-439/10-1, Es-85/8-1, Kr-1241/3-1), the DAAD
(H.R.), the Grant Agency of the Czech Rep.\ (Pr.\ 202/01/0764; J.M.),
discussions with N.\ Plakida, A.\ Moskvin, R.R.P.\ Singh, and
crystallographic analysis by A.\ Teresiak are greatfully acknowledged.
\end{acknowledgments}


\begin{thebibliography}{99}
\bibitem[*]{helge} Corresp.\ author (E-mail: helge@maugre.ucdavis.edu)
\bibitem[**]{hu} Present address: II.\ Physikal.\ Inst,\,
Universit\"at zu K\"oln, Z\"ulpicher Str.\ 77, D-50937 K\"oln,
Germany.
\bibitem[\dag]{mal}On leave from: Inst.\ of Physics, ASCR, Na Slovance
2, CZ-18221 Prague, Czech Rep.
\bibitem{Rice} T.M.\ Rice {\it et al.} Europhys.\ Lett.\ {\bf 23}, 445
(1993).

\bibitem{Dagotto} E.\ Dagotto, Rep.\ Progr.\ Physics {\bf 62}, 1525
(1999).

\bibitem{Mueller} T.F.A.\ M\"uller {\it et al.}, Phys. Rev. B {\bf
57}, R12 655 (1998).


\bibitem{Uehara} M.\ Uehara {\it et al.}, J.\ Phys.\ Soc.\ Jpn.\ {\bf
65}, 2764 (1996).

\bibitem{Isobe} M.\ Isobe {\it et al.}, Phys.\ Rev.\ B {\bf 57}, 613
(1998).

\bibitem{Kiryukhin} V.\ Kiryukhin {\it et al.}, Phys.\ Rev.\ B {\bf
63}, 1444418 (2001).

\bibitem{Ruck} K. Ruck {\it et al.}, Mat. Res. Bull. {\bf 36}, 1995
(2001).

\bibitem{Azuma} M.\ Azuma {\it et al.}, Phys.\ Rev.\ Lett.\ {\bf 73},
3463 (1994).

\bibitem{Sekar} C.\ Sekar {\it et al.}, Physica C (accepted for
publication).

\bibitem{structure} Chr.\ Teske {\it et al.}, Z.\ Anorg.\ Allg.\
Chem.\ {\bf 370}, 135 (1969); R.\ Arpe {\it et al.}, ibid.\ {\bf 426},
1 (1976).

\bibitem{Hiroi} Z. Hiroi {\it et al.}, J. Solid State Chem. {\bf 95},
230 (1991).

\bibitem{Jung} H.\ Petersen, {\it et al.}, Rev.\ Sci.\ Instrum. {\bf
66}, 1 (1995); C.\ Jung {\it et al.}, SPIE {\bf 3150}, 148 (1997).

\bibitem{Nucker} N.\ N\"ucker {\it et al.}, Z.\ Phys.\ B {\bf 67}, 9
(1987).

\bibitem{Troger} L. Tr\"oger {\it et al.}, Phys.\ Rev.\ B {\bf 46},
3283 (1992).

\bibitem{Teo} Transitions into unoccupied Cu 4$s$ states are also
allowed, but show a reduced transition probability by a factor of 20
compared to Cu 3$d$ final states: B.K.\ Teo and P.A.\ Lee, J.\ Am.\
Chem.\ Soc.\ {\bf 101}, 2815 (1979)

\bibitem{Fink}
J.\ Fink {\it et al.}, 
J.\ Electr.\ Spec.\ Relat.\ Phen.\ {\bf 66}, 395 (1994).

\bibitem{Knupfer}
M.\ Knupfer {\it et al.}, Phys.\ Rev.\ B {\bf 55}, R7291 (1997).

\bibitem{Nucker-00}
N. N\"ucker {\it et al.}, Phys.\ Rev.\ B {\bf 62}, 14 384 (2000).

\bibitem{Koepernik99} 
K.\ Koepernik {\it et al.}, Phys.\ Rev.\ B {\bf 59}, 1743 (1999).

\bibitem{Perdew81} J.P.\ Perdew {\it et al.}, Phys.\ Rev.\ B {\bf 23},
5048 (1981).
\bibitem{Eschrig89} H.\ Eschrig, {\em Optimized LCAO Method and the
Electronic Structure of Extended Systems} (Springer, Berlin, 1989).
\bibitem{Neudert}R.\ Neudert {\it et al.}, Phys.\ Rev. B {\bf 62},
10752 (2000).
\bibitem{remspin}A more sophisticated mapping including also cyclic
four-spin and anisotropic two-spin interactions \cite{brehmer,citro},
both are affected by the buckling, is left for future work.
\bibitem{brehmer}S.\ Brehmer {\it et al.}, Phys.\ Rev.\ B {\bf 60},
329 (1999).
\bibitem{citro}R.\ Citro and E.\ Orignac, Phys.\ Rev.\ B {\bf 65},
134413 (2002).
\bibitem{capriotti}L.\ Capriotti {\it et al.}, Phys.\ Rev.\ {\bf 65},
092406 (2002), showed that for coupled isotropic ladders
($J=J_{\perp}=J_{\parallel}$) at $T=0$ a quantum phase transition from
a spin gap to a N\'eel phase occurs near strong interladder (IL)
coupling $J_{IL}=0.3J$.
\bibitem{remy} As it also has been estimated for the strongest
interchain exchange in the closely related single chain compound
Sr$_2$CuO$_3$.\cite{Rosner} This value rises to about 3 to 4 meV for
the isomorphic Ca$_2$CuO$_3$ due to the reduced interchain distance
\cite{Rosner} which is of interest in the present context.
\bibitem{Rosner}H.\ Rosner {\it et al.}, Phys.\ Rev.\ B {\bf 56}, 3402
(1997).
\bibitem{rem2}The value of this ferromagnetic contribution depends on
the onsite energy difference between the two involved O 2$p$ orbitals,
the direct ferromagnetic exchange interaction between O and Cu
orbitals, as well as on the Cu-O(1)-Cu bond angle.  To estimate the
total ferromagnetic contribution, we mention that in CuGeO$_3$ with a
nearly 99$^\circ$ Cu-O-Cu bond angle it amounts to $\approx$ 15
meV.\cite{rosi} In the corner-shared BaCu$_2$Si$_2$O$_7$ with a
Cu-O-Cu bond angle of 124$^\circ$ very close to our case, a total
exchange integral of 24 meV has been estimated\cite{porier} which is
in accord with our estimate given above.
\bibitem{rosi}H.\ Rosner {\it et al.}, Phys.\ Rev.\ B {\bf 63}, 073104
(2001).
\bibitem{porier}M.\ Porier {\it et al.}, Phys.\ Rev. B {\bf 66},
054402 (2002).
\end{thebibliography}
\end{document}